# A Lyapunov Formulation for Efficient Solution of the Poisson and Convection-Diffusion Equations by the Differential Quadrature Method

Wen Chen and Tingxiu Zhong

Mechanical Engineering Department, Shanghai Jiao Tong University, Shanghai 200030, P.R.China.

The mail address for Wen Chen for proofs is P.O.Box 9601BB, Shanghai Jiao Tong University, Shanghai 200030, P.R.China.

Tel: 0086-021-62812679

Fax: 0086-021-62820892; 0086-021-62829425

E-mail: chenwwhy@hotmail.com

# 1. INTRODUCTION

Civan and Sliepcevich [1, 2] suggested that special matrix solver should be developed to further reduce the computing effort in applying the differential quadrature (DQ) method for the Poisson and convection-diffusion equations. Therefore, the purpose of the present communication is to introduce and apply the Lyapunov formulation which can be solved much more efficiently than the Gaussian elimination method. Civan and Sliepcevich [2] first presented DQ approximate formulas in polynomial form for partial derivatives in tow-dimensional variable domain. For simplifying formulation effort, Chen et al. [3] proposed the compact matrix form of these DQ approximate formulas. In this study, by using these matrix approximate formulas, the DQ formulations for the Poisson and convection-diffusion equations can be expressed as the Lyapunov algebraic matrix equation. The formulation effort is simplified, and a simple and explicit matrix formulation is obtained. A variety of fast algorithms in the solution of the Lyapunov equation [4-6] can be successfully applied in the DQ analysis of these two-dimensional problems, and, thus, the computing effort can be greatly reduced. Finally, we also point out that the present reduction technique can be easily extended to the three-dimensional cases.

## 2. DQ APPROXIMATE FORMULAS IN MATRIX FORM

The details about the differential quadrature method see reference [2]. By analogy with the procedure incorporating boundary conditions in Civan and Sliepcevich [1, 2], the DQ weighting coefficient matrices are modified in advance by using the boundary conditions. For example, considering the Dirichlet and Neumann boundary conditions in the x-direction ($x \in [0,1]$)

$$\phi(0, y) = h \tag{1}$$

$$\frac{\partial \phi(1, y)}{\partial x} = q. \tag{2}$$

Eq. (2) can be approximated by

$$\sum_{j=1}^{N} A_{Nj} \phi_j = q. \tag{3}$$

The function values at boundary points can be expressed by the unknown interior point function values, namely,

$$\phi_N = \frac{1}{A_{NN}} \left( q - A_{N1} h - \sum_{j=2}^{N-1} A_{Nj} \phi_j \right). \tag{4}$$

Substituting equations (4) and (1) into the DQ formulations for the first and second derivatives, respectively, we have

$$\frac{\partial \bar{\phi}}{\partial x} = \bar{A}_x \bar{\phi} + \bar{a}_x \tag{5}$$

$$\frac{\partial^2 \bar{\phi}}{\partial x^2} = \bar{B}_x \bar{\phi} + \bar{b}_x, \tag{6}$$

where $\bar{\phi} = \{\phi_2, \phi_3, \ldots, \phi_{N-1}\}$, $\bar{a}_x$ and $\bar{b}_x$ are the constant vectors, $\bar{A}_x$ and $\bar{B}_x$ are the modified (N-2)×(N-2) weighting coefficient matrices for the 1st and 2nd order derivatives, respectively.

Chen et al. [3] presented the DQ approximate formulas in matrix form for the partial derivative of the function ψ(x,y) in two-dimensional domain:

$$\begin{aligned} \frac{\partial^2 \psi}{\partial x^2} &= \bar{B}_x \psi + B_{ox}, & \frac{\partial^2 \psi}{\partial y^2} &= \psi \bar{B}_y^T + B_{oy}^T, \\ \frac{\partial \psi}{\partial x} &= \bar{A}_x \psi + A_{ox}, & \frac{\partial \psi}{\partial y} &= \psi \bar{A}_y^T + A_{oy}^T \end{aligned}, \tag{7}$$

where the unknown ψ is a n×m rectangular matrix rather than a vector as in references [1, 2], n and m is the number of inner grid points along x- and y- directions, respectively. The superscript T means the transpose of the matrices. $A_{0x}$ and $B_{0x}$ are generated by stacking the corresponding constant vectors $\vec{a}_x$ and $\vec{b}_x$ in Eqs. (5) and (6). For example,

$$A_{0x} = \begin{bmatrix} a_{1x} & a_{1x} & \cdots & a_{1x} \\ a_{2x} & a_{2x} & \cdots & a_{2x} \\ \vdots & \vdots & \ddots & \vdots \\ a_{nx} & a_{nx} & \cdots & a_{nx} \end{bmatrix}_{n \times m}. \tag{8}$$

$A_{0y}$ and $B_{0y}$ can be obtained in a similar way. For higher order partial derivatives, there exist similar matrix approximate formulas.

## 3. COMPUTATIONAL REDUCTION

### 3.1 Formulations in the Lyapunov matrix equation form

The Poisson equations can be normalized as:

$$\frac{\partial^2 \varphi}{\partial x^2} + \beta^2 \frac{\partial^2 \varphi}{\partial y^2} + S = 0, \tag{9}$$

where x and y are the dimensionless Cartesian coordinates, namely x, y $\in$ [0, 1], $\beta$ denotes the aspect ratio, S is a given strength, $\varphi$ is the desired variable. For more details see reference [1].

Applying the DQ matrix approximate formulas (7), the DQ formulation for equation (9) is given by

$$\bar{B}_x \bar{\varphi} + \beta^2 \bar{\varphi} \bar{B}_y^T + H = 0, \tag{10}$$

where $\bar{\varphi}$, $\bar{B}_x$ and $\bar{B}_y$ are (n-2)×(n-2) rectangular matrix, $H = S + B_{0x} + B_{oy}^T$. Since the boundary conditions have been taken into account in the formulation of weighting coefficient matrices $\bar{B}_x$ and $\bar{B}_y$, no additional equations are more required.

The equation governing steady-state convection-diffusion (e.g., equation (24) in reference [2] neglecting time derivative term) can be simplified as

$$\alpha \frac{\partial \phi}{\partial x} + \beta \frac{\partial \phi}{\partial y} = \frac{\phi}{4\alpha}, \tag{11}$$

where φ is the desired values as defined in equation (23) in reference [2], α and β are constants. In terms of the DQ matrix approximate formulas [7], we have

$$\alpha \bar{B}_x \phi + \beta \phi \bar{B}_y^T - \frac{1}{4a} \phi = Q, \tag{12}$$

where Q is constant matrix generated from the modified DQ weighting coefficient matrices as in Eq. (7). Furthermore, the above equation can be restated as

$$\left( \alpha \bar{B}_x - \frac{1}{4a} I \right) \phi + \beta \phi \bar{B}_y^T = Q. \tag{13}$$

The above DQ formulations (10) and (13) are the Lyapunov algebraic matrix equation. Compared with the conventional polynomial form in references [1, 2], they have more explicit matrix form.

### 3.2 Fast algorithms in the solution of the Lyapunov equations

Several efficient methods for solving the Lyapunov algebraic matrix equations have been developed in references [4-6]. To simplify the presentation, BS, HS and R-THR denote the methods proposed, respectively, by Bartels and Stewart [4], Golur, Nash and Loan [5], and Gui [6]. All these methods are stable and accurate. Considering the Lyapunov matrix equation

$$GX + XR = Q, \tag{14}$$

where G, R and Q are n×n, m×m and n×m constant matrices. The solution procedures in the BS, HS and R-THR methods generally include the following four steps.

**Step 1**: Reduce G and R into certain simple form via the similarity transformations $G^*=U^{-1}GU$ and $R^*=V^{-1}RV$.

**Step 2**: $F=U^{-1}QV$ for the solution of F.

**Step 3**: Solve the transformed equation $G^*Y+YR^*=F$ for Y.

**Step 4**: $X=UYV^{-1}$.

The respective computational effort is listed in table I. The total computing effort in these methods is $O(n^3+m^3)$ scalar multiplications. For the details on these methods see the corresponding references.

As can be seen from table I, the R-THR method requires $n^3+\frac{4}{3}m^3+7n^2m+5mn^2+n^2$ (or $14\frac{1}{3}n^3+n^2$ when n=m) scalar multiplications, and may be the most efficient in the solution of the Lyapunov matrix equations. By using the R-THR method, the same examples given in references [1] are recalculated by the DQ method, and the accuracies of results are coincident with those given by Civan and Sliepcevich [1]. However, the conventional approach in reference [1] required solving a linear simultaneous equations of $(N_x-2)(N_y-2)$ order by using the Gaussian elimination method, where $N_x$ and $N_y$ are the number of gird points along x- and y- directions, respectively. If $N=N_x=N_y$, about $\frac{1}{3}(N-2)^6$ multiplications were performed. In contrast, the present reduction approach requires about $14\frac{1}{3}(N-2)^3$ multiplications. Thus, the computational effort is only about 34% in using 7×7 grid points and 6% in using 11×11 grid points as much as that in reference [1]. The steady-state convection-diffusion (example 1 in reference [2]) is also computed by using the present technique, and the same computing reduction is achieved. The computational effort in the present DQ method for these cases is reduced in proportional to $(N-2)^3$. Reference [6] also pointed out that the parallel computation was very efficient in the solution of the Lyapunov equations.

The weighting coefficient matrices in the DQ method were found to be the centrosymmetric and skew centrosymmetric matrices if a symmetric grid spacing is used. The present authors [3] applied the factorization properties of centrosymmetric matrix to reduce computing effort by 75% in the DQ analysis of structural components. In the present cases with symmetric boundary conditions, we can factorize the centrosymmetric coefficient matrix $\bar{B}_x$ and $\bar{B}_y$ into two smaller size sub-matrices, nearly half, in all the four computing steps of the BS, HS and R-THR methods. Therefore, the computing effort can be further reduced to 8.5% under 7×7 grid points and 1.5% under 11×11 grid points as that in reference [1, 2]. The detailed discussions on the centrosymmetric matrix see reference [7] and are not presented here for the sake of brevity.

### 3.3 On the three-dimensional problems

For three-dimensional cases, we first convert into it into a set of ordinary differential equations by using the DQ matrix approximate formulas (7) and the Kronecker product. It is straightforward that the DQ matrix approximate formula for a set of ordinary differential equations is similar to formulas (7). Thus, the ordinary differential equations can be formulated into a Lyapunov matrix equation. The following examples can illustrate our idea more clearly. Considering the three-dimensional steady-state convection-diffusion equation (equation (52) in reference [2] neglecting time derivative term)

$$\frac{\partial c}{\partial x} = \beta \frac{\partial^2 c}{\partial y^2} + \gamma \frac{\partial^2 c}{\partial z^2}. \tag{15}$$

First, in terms of the DQ matrix approximate formulas (7), the above equation can be approximated as the following ordinary equations,

$$\overline{A}_x \mathcal{C} - \beta \mathcal{C} \overline{B}_y^T + Q = \gamma \frac{d^2 \mathcal{C}}{dz^2}, \tag{16}$$

where $\mathcal{C}$ is a $(N_x-2) \times (N_y-2)$ rectangular matrix, $Q = A_{0x} - \beta B_{0y}^T$. By using the Kronecker product of matrices [8], we have

$$\gamma \frac{d^2 \vec{C}}{dz^2} = \left[ \overline{A}_x \otimes I_y - \beta I_x \otimes \overline{B}_y \right] \vec{C} + Q, \tag{17}$$

where $\vec{C}$ is a $((N_x-2)(N_y-2)) \times 1$ vector stacked from matrix C, $\otimes$ denotes the Kronecker product. The DQ analog equation for the above ordinary differential equations can be written as

$$\left[ \overline{A}_x \otimes I_y - \beta I_x \otimes \overline{B}_y \right] \hat{C} - \gamma \hat{C} \overline{B}_z^T = R, \tag{18}$$

where $\hat{C}$ is a $((N_x-2)N_y-2)) \times (N_z-2)$ rectangular matrix, $R = \gamma B_{oz}^T - Q$. The equation is also a Lyapunov matrix equation. Thus, the reduction technique for the Lyapunov equation can be used to achieve a considerable savings in computational effort.

### 3.4. On time-dependent problems

The governing equation for transient-state convection-diffusion problems can be generally stated as (equation (24) in reference [2])

$$\frac{\partial \phi}{\partial t} + \frac{\phi}{4\alpha} = \alpha \frac{\partial \phi}{\partial x} + \beta \frac{\partial \phi}{\partial y}. \tag{19}$$

Similar to steady-state cases, the DQ analog equation can be expressed as

$$\frac{d\phi}{dt} = \left(\alpha \bar{B}_x - \frac{1}{4a}I\right)\phi + \beta\phi\bar{B}_y^T - Q. \tag{20}$$

The above DQ approximation equation is the simplified Riccati differential equation, namely, the quadratic nonlinear term in the Riccati differential equation is omitted. Choi and Laub [9] have successfully applied the fast algorithms for solving the Lyapunov equation to time-varying Riccati differential equation, while it is an easier task to apply these fast algorithms for calculating equation (20) in the same way as in reference [9]. The computational effort is reduced by three orders of magnitude as in the foregoing steady cases. Therefore, the extension of the present reduction DQ method to the transient convection-diffusion equations are also obviously applicable. For the details see reference [9].

## CONCLUSIONS

Compared with the Galerkin, Control-volume and finite difference methods, the differential quadrature method has proved to be a most efficient numerical technique in the calculation of the Poisson and convection-diffusion equations [1, 2]. The present work further minimizes the computational effort in the DQ solution of these cases. The principal advantages of the matrix approximate formulas are to offer a more compact and convenient procedure for obtaining an explicit matrix formulation and make the DQ method more efficient computationally for multi-dimensional problems by means of the existing techniques in the solution of the Lyapunov equations.

Acknowledgment –We are grateful to an referee whose thorough reviews were very helpful in the revision of the original manuscript of the present paper.

**REFERENCES**

1. F. Civan and C. M. Sliepcevich, Application of differential quadrature to transport processes, Int. J. Numer. Methods Engrg., 19, 711-724 (1983).

2. F. Civan and C. M. Sliepcevich, Differential quadrature method for multidimensional problems, J. Math. Anal. Appl., 101, 423-443 (1984).

3. Wen Chen, X. Wang and Y. Yu, Reducing the computational effort of the differential quadrature method, Numer. Methods for PDEs, 12, 565-577 (1996).

4. R. H. Bartels and G. W. Stewart, A solution of the equation AX+XB=C, Commun. AM., 15, 820-826, (1972).

5. G. H. Golub, S. Nash and C. Vanloan, A Hessenberg-Schur method for the problem AX+XB=C, IEEE Trans. Auto. Control, AC-24(6), 909-913 (1979).

6. B. Gui, A numerical solution of Lyapunov matrix equation and a partly parallel treatment (in Chinese), J. of Nanjing Univ. of Aeronautics and Astronautics, 24, 449-455 (1992).

7. L. Datta and S. D. Morgera, On the reducibility of centrosymmetric matrices application in engineering problems, Circuits Systems Signal Process, 8, 71-96 (1989).

8. P. Lancaster and M. Timenetsky, The Theory of Matrices with Applications, 2nd edition (Academic Press, Orlando, 1985).

9. C. H. Choi and A. J. Laub, Efficient matrix-valued algorithms for solving stiff Riccati differential equations, IEEE Trans. Auto. Contr., 35, 770-776 (1990).